\documentclass[
 reprint,amsmath,amssymb,aps,
]{revtex4-1}

\usepackage{graphicx}
\usepackage{latexsym}
\usepackage{amsfonts,amsmath,amssymb}

\usepackage{natbib}

\newcommand{\pr}{\operatorname{Pr}}
\newcommand{\po}{\operatorname{Po}}

\begin{document}

\title{Bayesian inference of time varying parameters in autoregressive
processes}

\author{Christoph Mark}
\affiliation{Friedrich-Alexander University Erlangen-N\"urnberg, Department of Physics, Biophysics Group}
\email[]{Contact: christoph.mark@fau.de}
 
\author{Claus Metzner}
\affiliation{Friedrich-Alexander University Erlangen-N\"urnberg, Department of Physics, Biophysics Group}

\author{Ben Fabry}
\affiliation{Friedrich-Alexander University Erlangen-N\"urnberg, Department of Physics, Biophysics Group}

\bibliographystyle{apsrev4-1}

\begin{abstract}
In the autoregressive process of first order AR(1), a homogeneous
correlated time series $u_t$ is recursively constructed as
$u_t = q\; u_{t-1} + \sigma \;\epsilon_t$, using random Gaussian
deviates $\epsilon_t$ and fixed values for the correlation coefficient
$q$ and for the noise amplitude $\sigma$. To model temporally
heterogeneous time series, the coefficients $q_t$ and $\sigma_t$ can be
regarded as time-dependent variables by themselves, leading to the
time-varying autoregressive processes TVAR(1). We assume here that the
time series $u_t$ is known and attempt to infer the temporal evolution
of the 'superstatistical' parameters $q_t$ and $\sigma_t$. We present a
sequential Bayesian method of inference, which is conceptually related
to the Hidden Markov model, but takes into account the direct
statistical dependence of successively measured variables $u_t$. The
method requires almost no prior knowledge about the temporal dynamics of
$q_t$ and $\sigma_t$ and can handle gradual and abrupt changes of these
superparameters simultaneously. We compare our method with a Maximum
Likelihood estimate based on a sliding window and show that it is
superior for a wide range of window sizes.

\end{abstract}

\maketitle

\section{Introduction}

The autoregressive process of first order, recursively defined as $u_t = q \; u_{t-1} + \sigma \; \epsilon_t$, represents the simplest model of a correlated time series and is therefore being used in many scientific or economic applications. A typical application in the natural sciences is the modeling of discrete time random walks in real space, where the random variable $u_t$ corresponds to the (vectorial) velocity of a particle at time $t$. In this case, the correlation coefficient $q$ is a measure of directional persistence, ranging from antipersistent behavior at $q=-1$, over non-persistent behavior at $q=0$, to persistent behavior at $q=1$. The noise amplitude $\sigma \geq 0$ then corresponds to the diffusivity of the particle.

In the original form of the AR(1) process, the model parameters $q$ and $\sigma$ are constants, thus assuming temporal homogeneity of the underlying random process. However, over sufficiently long time scales, this assumption is usually violated in most real world cases. For example, considering the diffusion of a test particle in a liquid, the temperature of the liquid might vary spatially, leading to an effective temporal modulation of $\sigma$ as the particle enters different regions. It is well-known that temporal heterogeneity of a correlated random walk can lead to anomalous features, such as a non-exponential decay of the velocity autocorrelation, or a non-Gaussian distribution of displacements within a given time interval. In physics, the explanation of anomalous statistical features in heterogeneous random process by a superposition of locally homogeneous processes has recently been termed 'superstatistics' \cite{Beck_2003,Beck_2005,Van_der_Straeten_2009,Beck_2011}.

The natural extension of the AR(1) process to heterogeneous situations is the time varying autoregressive process of first order, denoted as TVAR(1). In that case, the model parameters (or 'superparameters') $q_t$ and $\sigma_t$ have their own, deterministic or stochastic dynamics. This dynamics is either known (for example, if the spatial temperature profile of the liquid with the diffusing particle is experimentally controlled), or it has to be inferred from the measured time series $u_t$ alone. 

In this paper, we are interested in the latter case and present a method to extract the temporal evolution of $q_t$ and $\sigma_t$ directly from $u_t$. Based on sequential Bayesian updating, the method can be applied with only very limited prior assumptions about the dynamical behavior of these superparameters, yet allows for incorporating dynamical models of the superstatistical process if available. Importantly for many applications, the method is able to detect sudden changes and slow dynamics of $q_t$ and $\sigma_t$ simultaneously.

Extracting the temporal evolution of the superparameters can be even more revealing about a complex system than the knowledge of the direct time series $u_t$. For example, in cases where $u_t$ describes the behavior of an agent, temporal variations of the superparameters may reflect changes in the environment of the agent, or changes in its internal state.

Similar inference problems have been tackled by a variety of methods, including the conceptually simple sliding window analysis \cite{Zivot_2006,Sun_2011}, as well as more advanced approaches, such as \emph{recursive least squares} \cite{Hayes_1996} and basis function approaches \cite{Hall_1977}.

In this paper, we compare our proposed method to the maximum likelihood estimation within a sliding window, and evaluate the quality of parameter extraction using simulated trajectories with known time traces of $q_t$ and $\sigma_t$.

\section{Method}
\subsection{TVAR(1) and sequential Bayesian updating}
We consider the measured time series $\{u_t\}$ of length $N$, with $t=0,...,N-1$. The observations are connected via the time-varying parameters $\{q_t, \sigma_t\}$ ($t=1,...,N-1$). The recursive relation is given by the TVAR(1) process:
\begin{align} \label{eqn:TVAR}
u_t = q_t \; u_{t-1} + \sigma_t \; \epsilon_t~,
\end{align}
where $\epsilon_t$ denotes the noise term and is drawn from a standard normal distribution.

Our proposed method can be applied to the general case, where each measurement $u_t$ is a vector, and $q_t$ and $\sigma_t$ are matrices. However, in this work we are particularly interested in the special case where $u_t$ is a velocity vector of a particle, $q_t$ and $\sigma_t$ are scalars, and the components of the noise term are assumed to be $iid$. Note that these restrictions imply local isotropy of the random process.

In order to infer the values of the superparameters of the TVAR(1)-process, we need to state the likelihood function, denoted by $L$, describing the probability of the measured values $\{u_t\}$, given the parameter values $\{q_t, \sigma_t\}$. In the case of the first-order process discussed here, the likelihood can be factorized into 'one-step' terms, since the current value $u_t$ only depends on the previous value $u_{t-1}$, apart from the model parameters:
\begin{align}
L(\{u_t\}) \equiv p(\{u_t\}|\{q_t, \sigma_t\}) = p(u_0) \prod_{t=1}^N p(u_t|q_t, \sigma_t,u_{t-1})~,
\end{align}
where $p(u_0)$ can be seen as a constant, as it plays no further role in the inference process. The one-step-likelihood follows directly from Eq. (\ref{eqn:TVAR}):
\begin{align} \label{eqn:likelihood}
L(u_t;u_{t-1}) &\equiv p(u_t|q_t, \sigma_t; u_{t-1}) = \nonumber\\
&= \frac{1}{(2 \pi \sigma_t^2)^{m/2}} \exp\left({- \frac{\left(u_t - q_t u_{t-1}\right)^2}{2 \sigma_t^2}}\right)~,
\end{align}
where $m$ is the dimensionality (number of vector components) of $u_t$. This term provides a model-specific description of the direct correlation between observed (measured) values. Note that at time $t$, the previous measurement $u_{t-1}$ is known. Therefore, it is not treated as a random variable but rather as a constant. In the notation above, this constant is separated by a semicolon. The Bayesian inference scheme devised below is based on this time-dependent likelihood term and can thus be seen as an extension to the Hidden Markov Model (HMM), in which all observations are assumed to be independent \cite{Rabiner_1989}.

In the general Bayesian framework, the parameter estimates are expressed by their joint posterior distribution $\po\left(\{q_t, \sigma_t\}\right) \equiv p(\{q_t, \sigma_t\}|\{u_t\})$, which is gained by multiplying the likelihood with a prior distribution $\pr\left(\{q_t, \sigma_t\}\right) \equiv p(\{q_t, \sigma_t\})$, reflecting our knowledge of the parameters before seeing the data:
\begin{align}
\po\left(\{q_t, \sigma_t\}\right) \propto L\left(\{u_t\}\right)\;\pr\left(\{q_t, \sigma_t\}\right)
\end{align}
While computing this potentially very high dimensional posterior distribution is feasible using approximate Markov Chain Monte Carlo methods, the problem of finding an appropriate prior that is flexible enough to account for sudden changes as well as slow parameter dynamics remains. In order to evade this difficulty, we propose an iterative inference algorithm, which incorporates the different aspects of the parameter dynamics at every time step.

\begin{figure}[h!]
\begin{center}
\includegraphics[width=0.9\columnwidth]{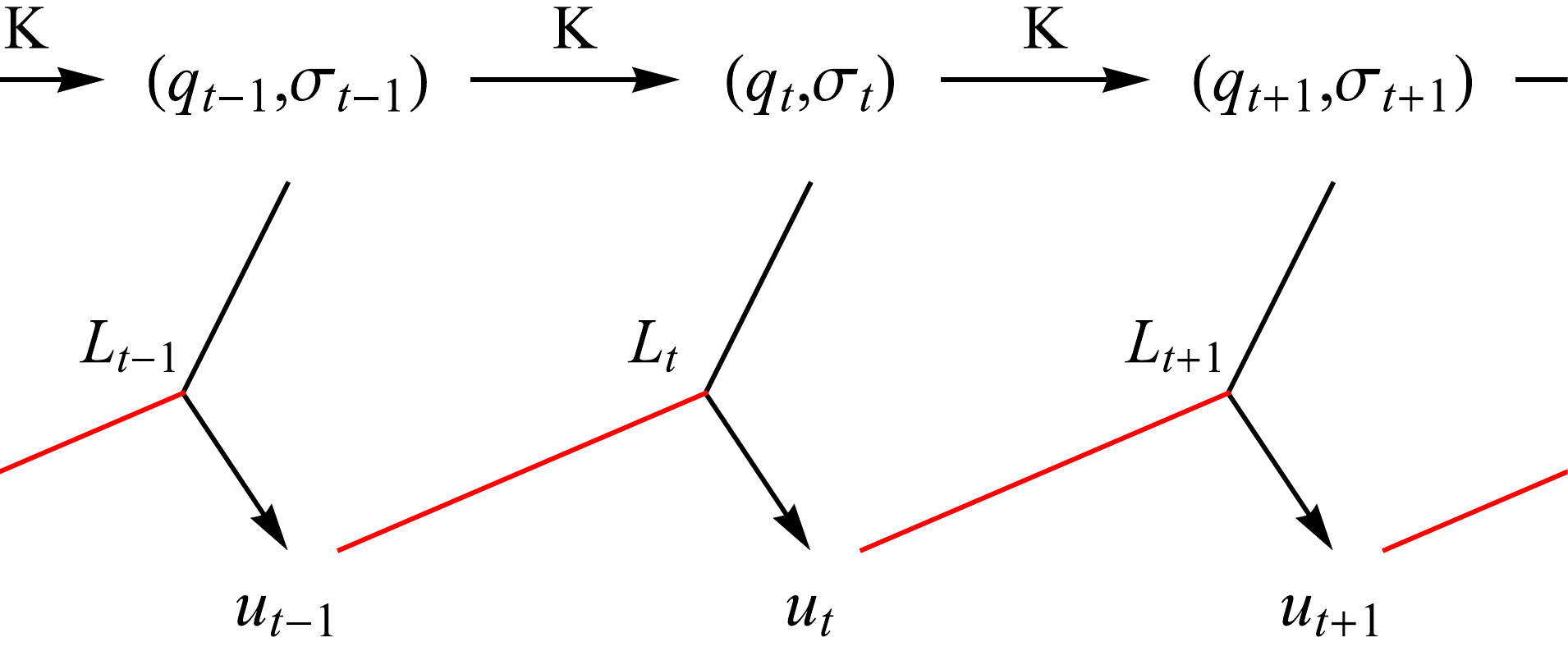}
\caption{\label{fig:HMM}
Illustration of the conceptual relation of the proposed model to the known Hidden Markov Model. The direct statistical dependence of the measured values is depicted in red.}
\end{center}
\end{figure}

Using the one-step likelihood term, the Bayes scheme can be stated for a single time step, given an appropriate prior for the parameters at time $t$:
\begin{align} \label{eqn:onestepbayes}
\po\left(q_t, \sigma_t\right) \propto L\left(u_t;u_{t-1}\right)\;\pr\left(q_t, \sigma_t\right) ~.
\end{align}
While the likelihood function in Eq. (\ref{eqn:likelihood}) describes the relation of subsequent \emph{observed} values, the prior distribution $\pr(q_t, \sigma_t) \equiv p(q_t, \sigma_t)$ represents our belief in the parameter values at time $t$ before seeing the corresponding data point $u_{t}$, and allows us to incorporate the expected temporal behavior of the latent superparameters. This is done using a transformation $K$, which relates the computed posterior distribution of the previous time step with the current prior distribution:
\begin{align} \label{eqn:prior}
\pr(q_t,\sigma_t) = K\left(\;\po(q_{t-1},\sigma_{t-1})\;\right)~.
\end{align}
The structure of the proposed model can be directly compared to the known HMM. This analogy is illustrated in Fig. \ref{fig:HMM}, where $K$ takes the role of the state transition matrix, while $L_t$ can be interpreted as a generalized observation matrix with the important difference that it also takes into account the previous observation (red line).

In practice, there is often only little knowledge about the underlying dynamics of the superstatistical parameters $q_t$, $\sigma_t$. It is therefore crucial to find a transformation $K$ which on the one hand keeps the restrictions on the evolution of these time-varying parameters as small as possible and on the other hand minimizes the estimation error. Here, we propose a two-step transformation $K$ to form the current prior distribution from the previous posterior:
\begin{align}
K = K_2 \circ K_1 ~,
\end{align}
where the first transformation introduces a minimal probability for the current value of the parameter-tuple,
\begin{align} \label{eqn:pmin}
K_1 ~~:~~ p(q_t,\sigma_t) ~\longrightarrow~ Max\left[p_{\text{min}},p(q_t,\sigma_t)\right]~.
\end{align}
The mapping ensures that there is always a small probability even for parameter values that deviate strongly from the preceding ones, allowing the method to detect abrupt changes of the correlation coefficient and/or noise amplitude.

The second transformation $K_2$ describes a convolution of a probability distribution with a box kernel, denoted $B$:
\begin{align} \label{eqn:kernel}
K_2(~p(q_t,\sigma_t)~) =  \left( B \; \ast \; p\right)(q_t,\sigma_t) ~,
\end{align}
where $(~\ast~)$ denotes the convolution and the two-dimensional box kernel is defined as
\begin{align}
B(x_1, x_2) = \Theta\left(R - |x_1|\right)\;\Theta\left(R - |x_2|\right) ~.
\end{align}
Here, $R$ is the radius of the kernel and $\Theta(x)$ is the Heaviside step function. This transformation can be interpreted as a moving average filter, \emph{blurring} the probability distribution given as the argument. Applied to the posterior distribution of $(q_{t-1},\sigma_{t-1})$, the transformation $K_2$ thus allows for an accurate detection of slow parameter dynamics. It is important to note that the joint posterior distribution of the parameters is normalized at every time step, since the mapping $K$ does not preserve normalization.

\begin{figure}[h!]
\begin{center}
\includegraphics[width=1.0\columnwidth]{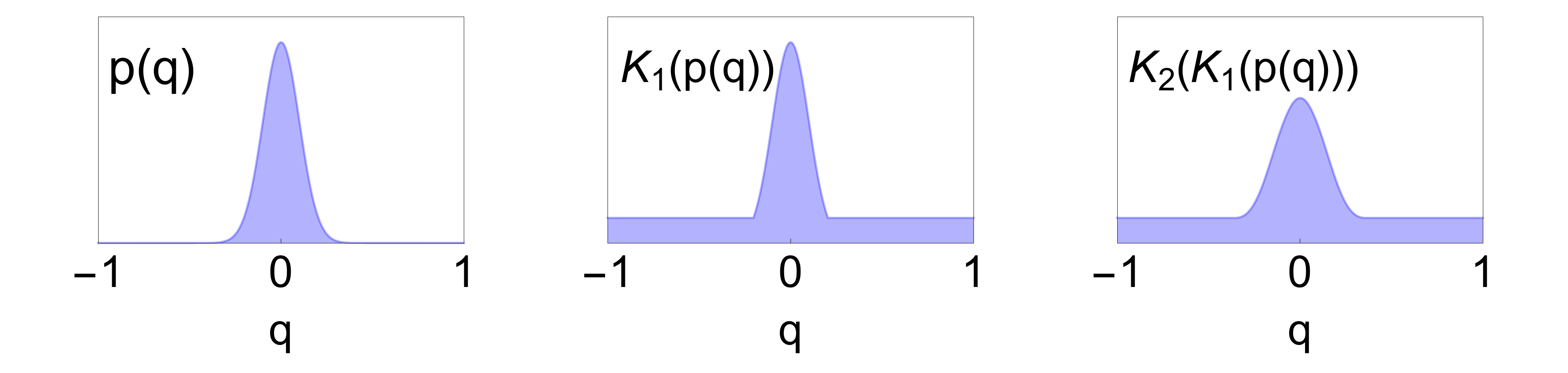}
\caption{\label{fig:kernel_illu}
Transformation of a probability distribution using the mapping $K$. Here, $K_1$ introduces a minimal probability of realizing any parameter value, while $K_2$ ''blurs'' the distribution.}
\end{center}
\end{figure}

Figure \ref{fig:kernel_illu} shows the effect of the transformations $K_1$ and $K_2$ on a generic probability distribution. For presentation reasons, we only show the transformation of an univariate distribution.

\subsection{Bi-directional inference}
The method described above infers the time-varying superstatistical parameters in an iterative way, moving forward along the time axis. The latent time series, $\{q_t, \sigma_t\}$, thus inherits a property from the measured values, $\{u_t\}$, that is characteristic for the latter: the current value can only depend on past values, not on future ones. While this causality condition on the observable time series follows directly from Eq. (\ref{eqn:TVAR}), it imposes an unnecessary restriction on the latent time series -- and therefore on the parameters we want to estimate.

In contrast to the dynamics of $u_t$, described by the TVAR(1) process, the dynamics of the superparameters, described by the mapping $K$ proposed above, is reversible. Formally, this property is called detailed balance, c.f. \cite{Bhanot_1988}:
\begin{align}
&p\left( (q_t,\sigma_t) =i ~|~ (q_{t-1},\sigma_{t-1})=j \right) = \nonumber\\
&p\left( (q_t,\sigma_t)=j ~|~ (q_{t-1},\sigma_{t-1})=i \right)~.
\end{align}
The equation above holds for $K$, since the proposed mapping is symmetric around the current parameter values, $(q_t, \sigma_t)$.

Here, we propose an inference algorithm that makes explicit use of the reversible dynamics of the superstatistical parameters, extending the strictly forward moving procedure described above.

\begin{figure}[h!]
\begin{center}
\includegraphics[width=0.9\columnwidth]{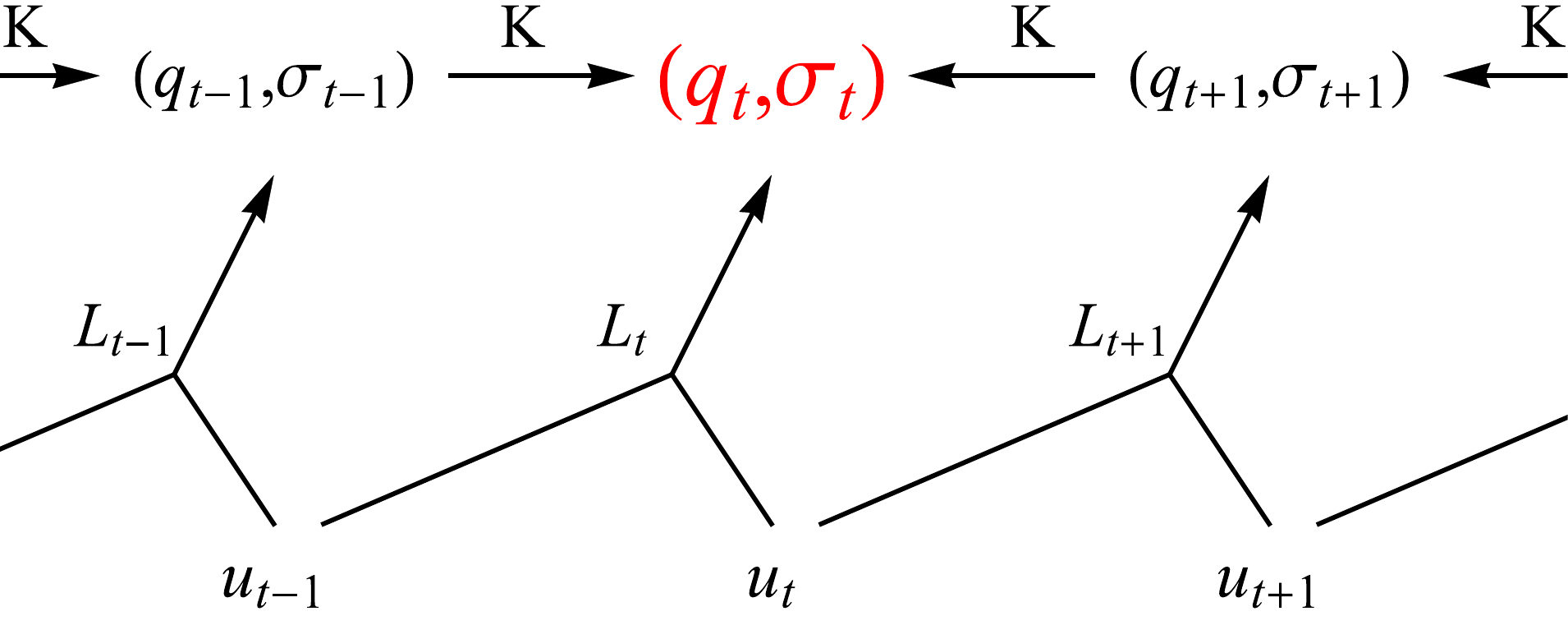}
\caption{\label{fig:bidirec}
Illustration of the iterative, bi-directional inference of the parameters $(q_t,\sigma_t)$.}
\end{center}
\end{figure}

In order to estimate the local persistence and noise amplitude at time $t$, we first need to compute a prior distribution. Moving forward in time, the prior is gained by applying $K$ on the previous posterior distribution, see Eq. (\ref{eqn:prior}):
\begin{align}
\pr^F\left(q_t,\sigma_t\right) &= K\left(\po\left(q_{t-1},\sigma_{t-1}\right)\Big|_{\genfrac{}{}{0pt}{}{q_{t-1}=q_t}{\sigma_{t-1}=\sigma_t}}\right) ~ ,
\end{align}
where the upper index $F$ indicates the direction in time, \emph{forward}. Note that for clarification, we denote the posterior as $p(q_{t-1}, \sigma_{t-1}| L_{t-1})$, which is equal to $p(q_{t-1}, \sigma_{t-1}| u_{t-1}, u_{t-2})$. Similarly, we can compute a prior of the parameters based on future values, by starting the inference process using the last likelihood, $L_{N-1}$, and moving backwards in time. It is important to note here that we use the same one-step likelihood functions as above, so that the observable values still only depend on past values. This \emph{backward}-prior is computed using the following iterative scheme:
\begin{align}
\pr^B\left(q_t,\sigma_t\right) &= K\left(\po\left(q_{t+1},\sigma_{t+1}\right)\Big|_{\genfrac{}{}{0pt}{}{q_{t-1}=q_t}{\sigma_{t-1}=\sigma_t}}\right) ~ .
\end{align}
Applying the one-directional sequential updating procedure in both directions, we now have two independent priors that can be combined with the likelihood at time $t$, in order to get the bi-directional posterior distribution of the superstatistical parameters:
\begin{align}
\po(q_t,\sigma_t) \propto \pr^F(q_t,\sigma_t)\;\pr^B(q_t,\sigma_t)\;L(u_t;u_{t-1}) ~,
\end{align}
which has to be properly normalized. The estimates of $q_t$ and $\sigma_t$ thus make use of all available data. While this improves estimation quality, it also inhibits an on-line implementation of the algorithm. The proposed inference scheme is illustrated in Fig. \ref{fig:bidirec}.

\subsection{Grid-based implementation} \label{sec:grid}
Since the iterative application of $K$ renders the analytic treatment of the posterior distributions intractable, we rely on a grid-based implementation of the described inference scheme. The parameter space, $(q,\sigma)$, is discretized in equally spaced points, resulting in a $N_q \times N_\sigma$-grid:
\begin{align}
 \begin{matrix}
  (q_1,\sigma_1) & (q_1,\sigma_2) & \cdots & (q_1,\sigma_{N_\sigma}) \\
  (q_2,\sigma_1) & (q_2,\sigma_2) & \cdots & (q_2,\sigma_{N_\sigma}) \\
  \vdots  & \vdots  & \ddots & \vdots  \\
  (q_{N_q},\sigma_1) & (q_{N_q},\sigma_2) & \cdots & (q_{N_q},\sigma_{N_\sigma})
 \end{matrix}
\end{align}
Subsequently, our current belief of the value of the superparameters can be described by a probability mass function, given as a $(N_q \times N_\sigma)$-dimensional matrix:
\begin{align}
\left( p\left(q_t,\sigma_t\right) \right)_{nm} = p\left(q_t = q_n,\sigma_t = \sigma_m\right) ~.
\end{align}
With the likelihood $L$ discretized in the same way, the computation of the posterior from the prior distribution and the likelihood reduces to a component-by-component multiplication of the corresponding matrices.

\subsection{Maximum Likelihood estimate within sliding window}
In order to assess the performance of the method proposed above, we provide a short discourse of a simple alternative method to analyze potentially heterogeneous time series, namely the sliding-window analysis based on a Maximum Likelihood estimation.

Assuming a homogeneous AR(1)-process for times $t'$ within the interval $I_t=\{t-w/2,...,t+w/2\}$, centered around some point in time $t$, with persistence $q_t$ and noise amplitude $\sigma_t$, one can state the log-likelihood for these parameters as follows:
\begin{align} \label{eqn:LLH}
&\log~ p\left( \left\{ u_{t'} \right\}_{t'\in I_t} |q_t,\sigma_t \right) ~\propto \nonumber\\
&\sum_{t'\in I_t} \left( - \frac{\left(u_{t'} - q_t u_{t'-1}\right)^2}{2 \sigma_t^2} - log \left( 2 \pi \sigma_t^2\right) \right) ~ .
\end{align}
Note that we can ignore the probability of observing $u_{t'= t - \frac{w}{2} - 1}$ because it does not affect the subsequent maximization of the log-likelihood function.

Maximizing Eq. (\ref{eqn:LLH}) with respect to $q_t$ and $\sigma_t$ yields the following Maximum Likelihood estimators:
\begin{align}
\hat{q}_t = \frac{\sum_{t'\in I_t} ~~~u_{t'}.u_{t'-1} }{\sum_{t'\in I_t} ~~u_{t'-1}.u_{t'-1} } ~ ,
\end{align}
where $(~.~)$ denotes the dot product. The estimator for the noise amplitude can be stated as
\begin{align}
\hat{\sigma}_t = \sqrt{\frac{1}{2 w} \left| \sum_{t'\in I_t} \left(u_{t'} - \hat{q}_t u_{t'-1}\right)^2 \right| } ~,
\end{align}
and depends on the estimate for $q_t$. In order to analyze a potentially heterogeneous time series using these estimators, one partitions the data into overlapping segments of length $w$ -- thus the term 'sliding window' -- and is able to estimate the \emph{local} persistence and noise amplitude.

Note that in the sliding window approach, one loses $w-1$ of the $q_t$ and $\sigma_t$ estimates, whereas the Bayesian method yields $N-1$ estimates for $N$ data points. This can be a significant advantage for small data sets.

\begin{figure}[h!]
\begin{center}
\includegraphics[width=0.9\columnwidth]{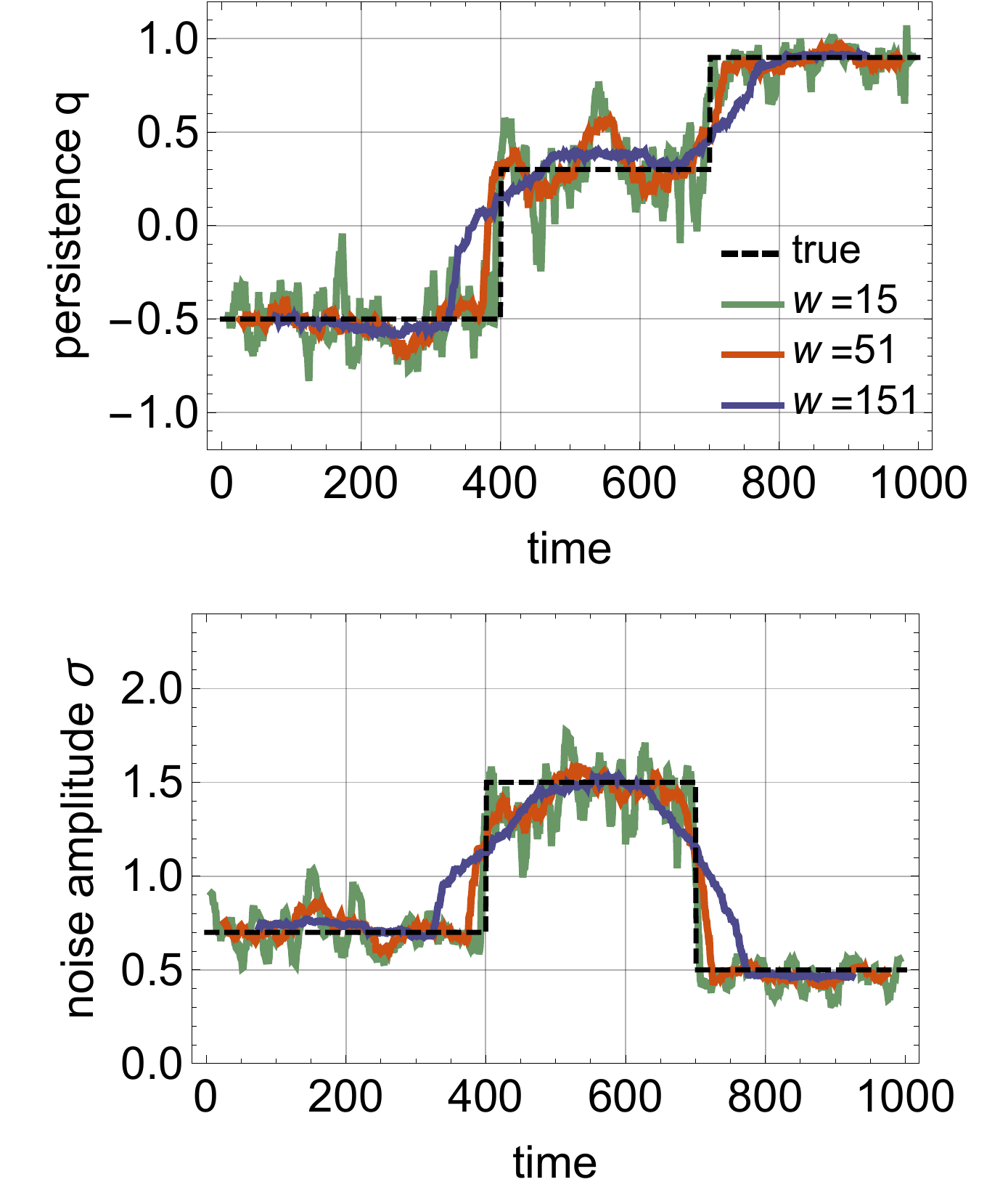}
\caption{\label{fig:MLrecon}
Estimated parameter values of the local persistence (top) and noise amplitude (bottom) for different window sizes, together with the true parameters values (dashed). Small window widths result in a fast detection of abrupt parameter changes but exhibit strong fluctuations. Large window sizes provide smoothly changing parameter estimates with only small noise, but are unable to detect sudden changes.}
\end{center}
\end{figure}

Furthermore, the assumption of constant parameters $q$ and $\sigma$ within each window of length $w$ represents a great weakness of the sliding window approach, since it cannot be fulfilled for truly heterogeneous time series. Finally, the choice of the 'window size', $w$, strongly affects the resulting reconstruction of the time-varying parameters (Fig. \ref{fig:MLrecon}).

\section{Results}
We subsequently assess the estimation quality of the proposed Bayesian inference scheme by applying the grid-based implementation, introduced in Sec. \ref{sec:grid}, to a number of simulated two-dimensional trajectories showing different temporal behavior of the superparameters.

Here, the parameter space $(q_t,\sigma_t)$ is discretized using a $200 \times 200$ dimensional quadratic grid, with the following boundaries:
\begin{align} \label{eqn:boundaries}
-1.5 < q_t < 1.5~, ~~~~~~ 0 < \sigma_t < 3 ~~ \forall t.
\end{align}
The minimal probability for reaching every point in the parameter space, $p_{min}$ (cf. Eq. (\ref{eqn:pmin})), is set to $p_{min}=10^{-7}$. In the discrete implementation of the algorithm, the box kernel used to cover slow parameter dynamics is chosen to be a $5\times 5$-matrix with equal values, $1/25$. The radius $R$, as defined in Eq. (\ref{eqn:kernel}) thus equals $R = 2 \delta$, with $\delta = 3/200$ being the distance between to adjacent points on the parameter grid.

\subsection{Regime-switching process}
\begin{figure}[h!]
\begin{center}
\includegraphics[width=0.9\columnwidth]{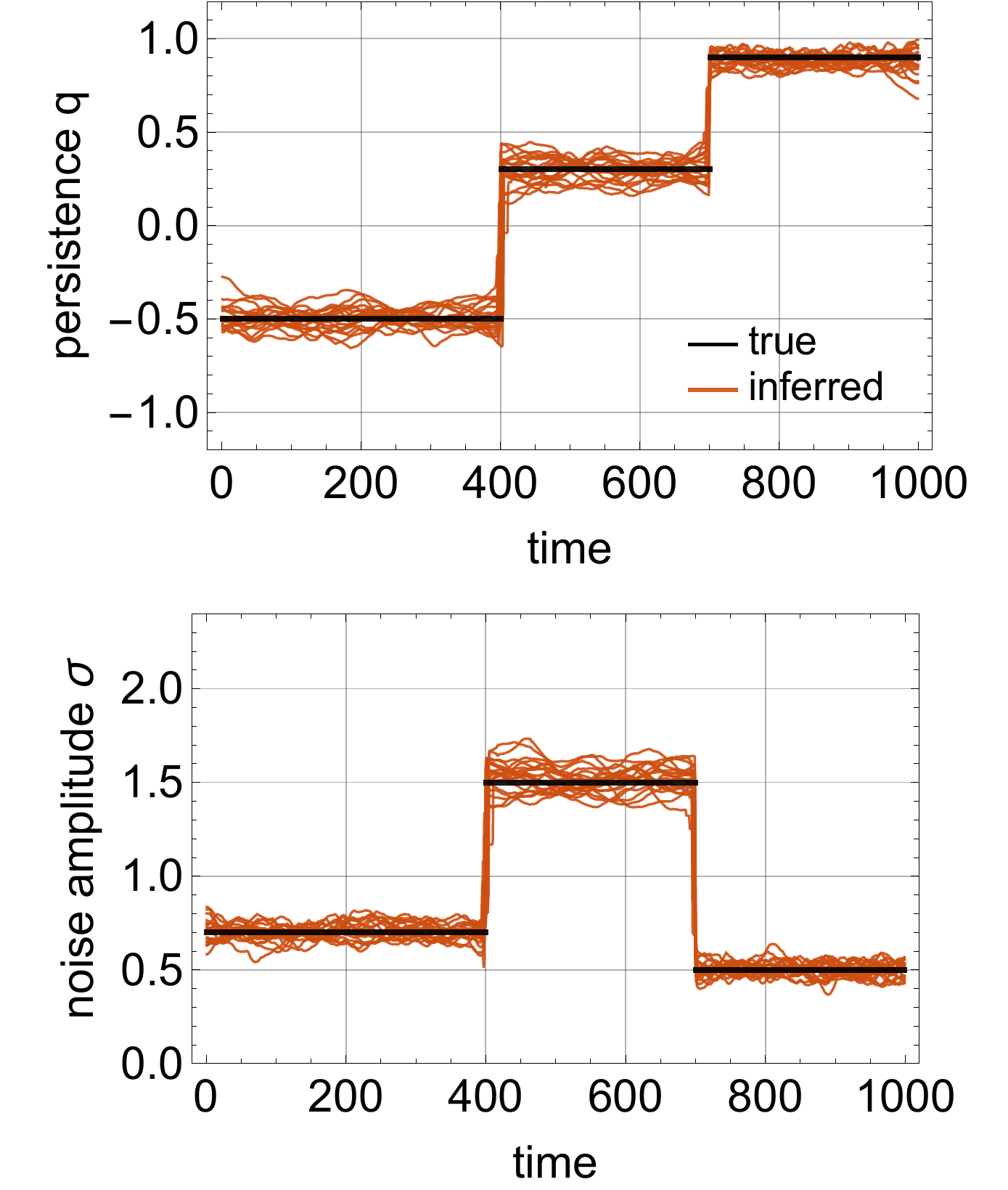}
\caption{\label{fig:regimerecon}
Inference of piecewise constant parameter values (black). Orange lines show inferred parameter values of persistence (top) and noise amplitude (bottom) for $20$ realizations of the TVAR(1) process using the true parameter values.}
\end{center}
\end{figure}

As a first demonstration of the proposed inference method, we simulate two-dimensional trajectories based on piecewise constant parameters. This so-called regime-switching process exhibits abrupt changes of both, persistence and noise amplitude. Here, the time-varying parameter values are chosen as follows:
\begin{align} \label{eq:regimevalues}
(q_t, \sigma_t) = \left\{
\begin{array}{l l l}
(-0.5,0.7)  & \textrm{for} &  0 < t \leq 400  \\
(0.3,1.5)  & \textrm{for} &  400 < t \leq 700  \\
(0.9,0.5)  & \textrm{for} &  700 < t \leq 1000  \\
\end{array}
\right.
\end{align}
Figure \ref{fig:regimerecon} shows the inferred parameter series of the persistence (top) and noise amplitude (bottom) for $20$ simulated trajectories, using the grid-based implementation of the bi-directional inference algorithm as described in Sec. \ref{sec:grid}. In this case, the algorithm produces an immediate response to the abrupt parameter changes. In contrast, using the sliding-window approach, the chosen window width limits the temporal response to an abrupt change of parameter values (see Fig. \ref{fig:MLrecon}).

\begin{figure}[h!]
\begin{center}
\includegraphics[width=0.9\columnwidth]{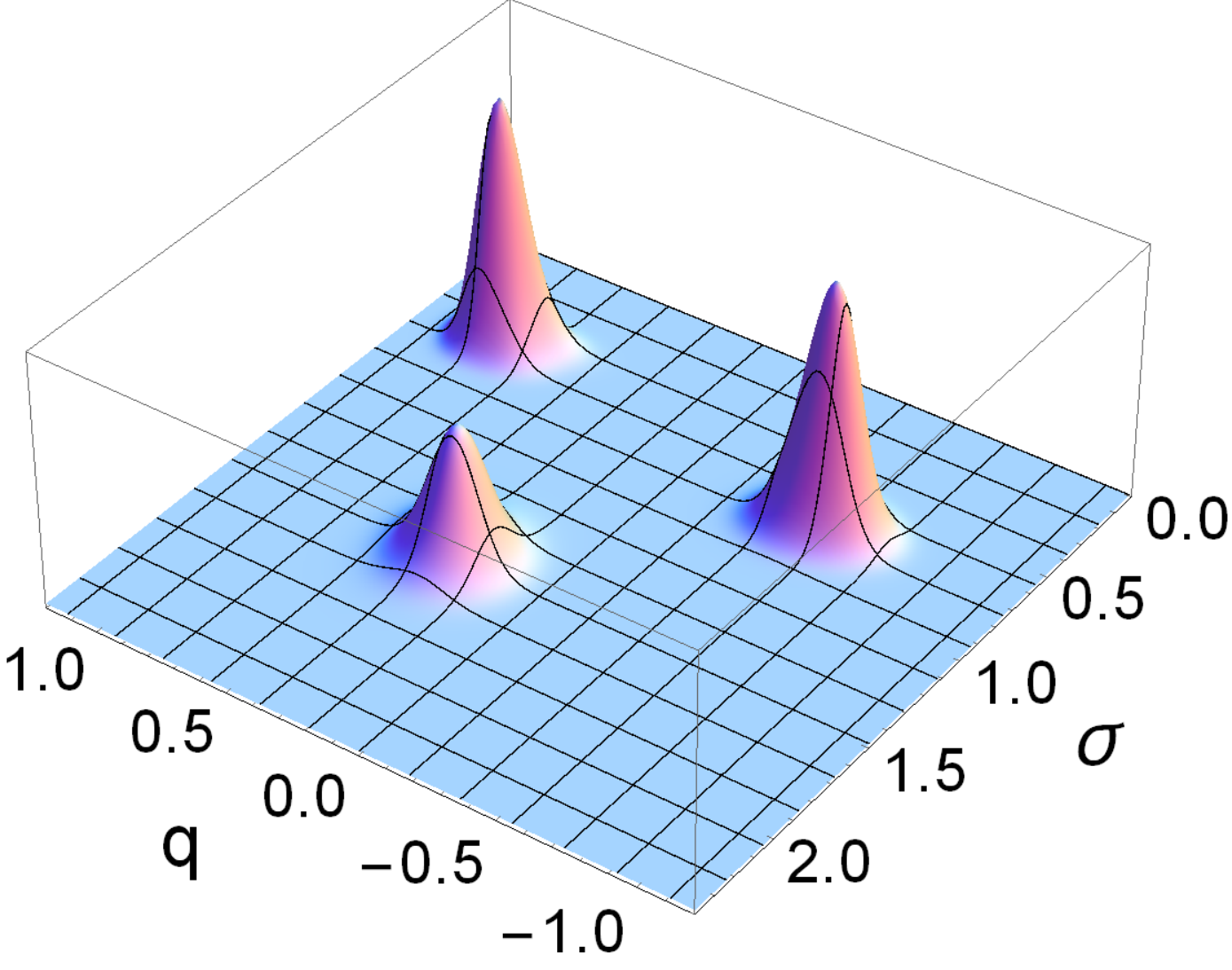}
\caption{\label{fig:regimedist}
Time-averaged posterior distribution of a single reconstructed parameter sequence with piecewise constant parameter values. The distribution shows the three distinct peaks coinciding with the parameter regimes defined in Eq. \ref{eq:regimevalues}, while the width of each peak incorporates the uncertainty of all estimates over time.}
\end{center}
\end{figure}

Since the Bayesian method naturally preserves the joint parameter distribution at every time step as a measure of how certain the estimates actually are, we can compute the time-averaged posterior distribution, $\left\langle \po(q_t,\sigma_t) \right\rangle_t$, for the analyzed trajectory. Figure \ref{fig:regimedist} shows the time-averaged posterior distribution corresponding to the estimated parameter values shown as a red line in Fig. \ref{fig:regimerecon}. In this specific case, three parameter regimes appear as clearly seperated. General cases with mutually merging regimes, corresponding to overlapping peaks in the probability distribution, can be described as well.

\subsection{Linearly changing parameter values}
\begin{figure}[h!]
\begin{center}
\includegraphics[width=0.9\columnwidth]{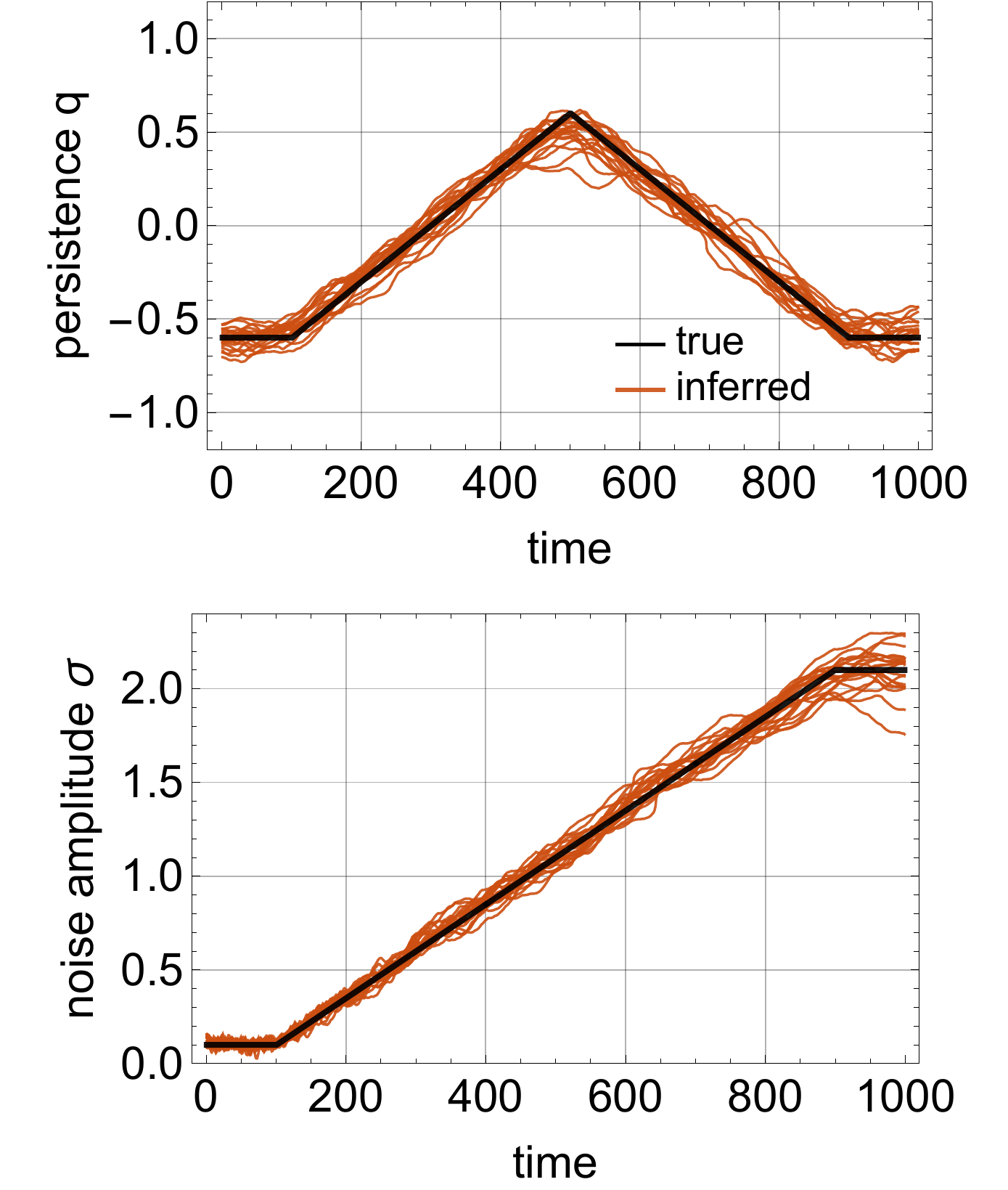}
\caption{\label{fig:linearrecon}
Inference of linearly changing parameter values (black). Orange lines show inferred parameter values of persistence (top) and noise amplitude (bottom) for $20$ realizations of the TVAR(1) process using the true parameters values.}
\end{center}
\end{figure}

\begin{figure}[ht]
\begin{center}
\includegraphics[width=0.9\columnwidth]{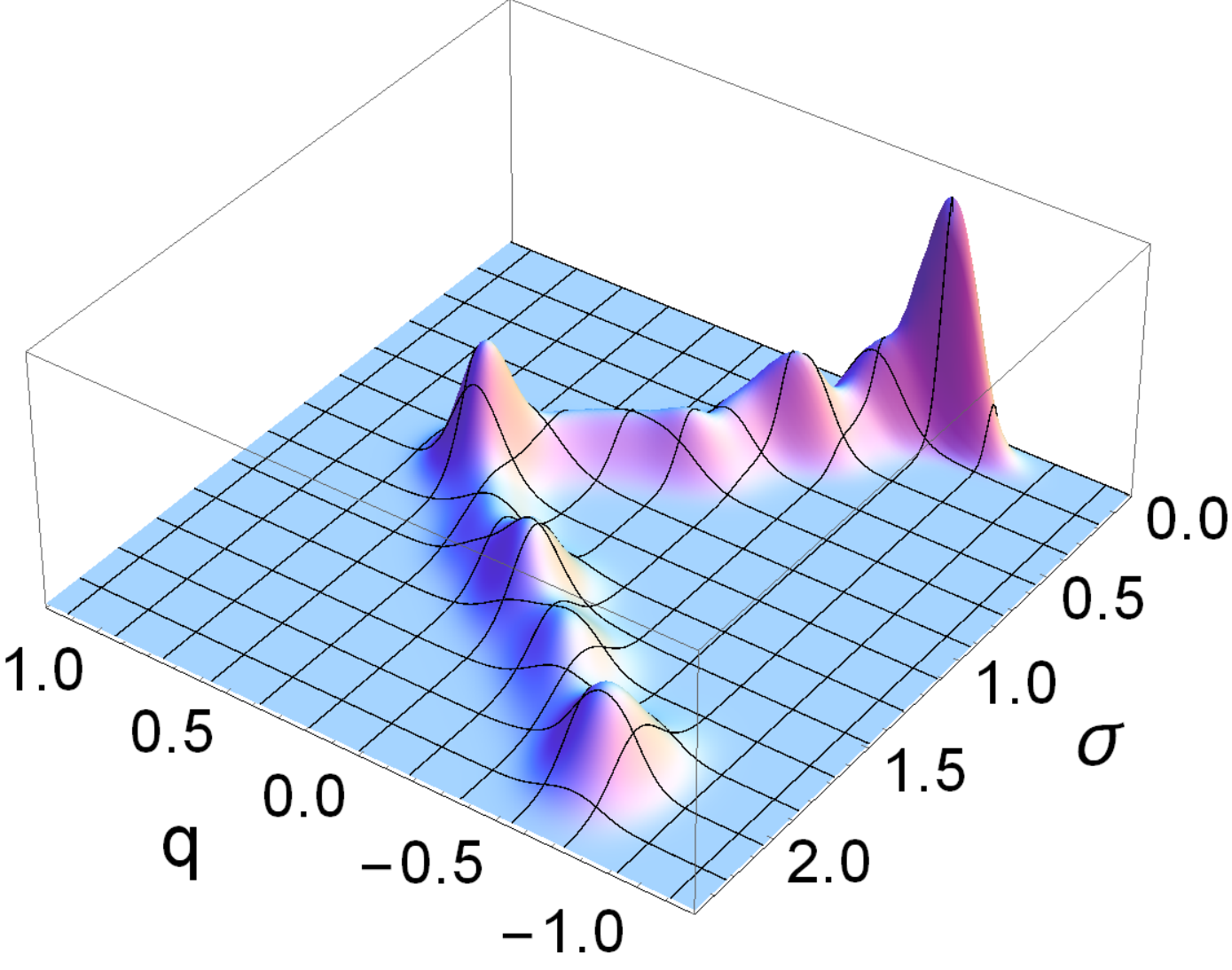}
\caption{\label{fig:lineardist}
Time-averaged posterior distribution of a single reconstructed parameter sequence with linearly varying parameter values. The distribution accurately reproduces the correlations between persistence $q_t$ and noise amplitude $\sigma_t$.}
\end{center}
\end{figure}

In contrast to the abruptly changing parameter values investigated above, this example shows the estimation of local persistence and noise amplitude values which change at a piecewise constant rate. The change in parameter value per time step, $\Delta q = q_{t+1}-q_t$ and $\Delta \sigma = \sigma_{t+1}-\sigma_t$, respectively, are chosen as follows:
\begin{align} \label{eq:linearvalues}
(\Delta q, \Delta \sigma) = \left\{
    \begin{array}{l l l}
    (0,0)  & \textrm{for} &  0 < t \leq 100  \\
    (0.003,0.0025)  & \textrm{for} &  100 < t \leq 500  \\
    (-0.003,0.0025)  & \textrm{for} &  500 < t \leq 900  \\
    (0,0)  & \textrm{for} &  900 < t \leq 1000  \\
    \end{array}
    \right.
\end{align}
The true parameter values (dashed) are shown alongside estimated ones (red and gray lines) for multiple realizations of the process in Fig. \ref{fig:linearrecon}. As for the regime-switching case shown above, we compute the time-averaged posterior distribution of a single series of reconstructed parameter values (corresponding to the red line in Fig. \ref{fig:linearrecon}). The resulting distribution, displayed in Figure \ref{fig:lineardist}, clearly captures the predefined correlations between $q_t$ and $\sigma_t$. 

\subsection{Sinusoidal parameter changes}
\begin{figure}[h!]
\begin{center}
\includegraphics[width=0.9\columnwidth]{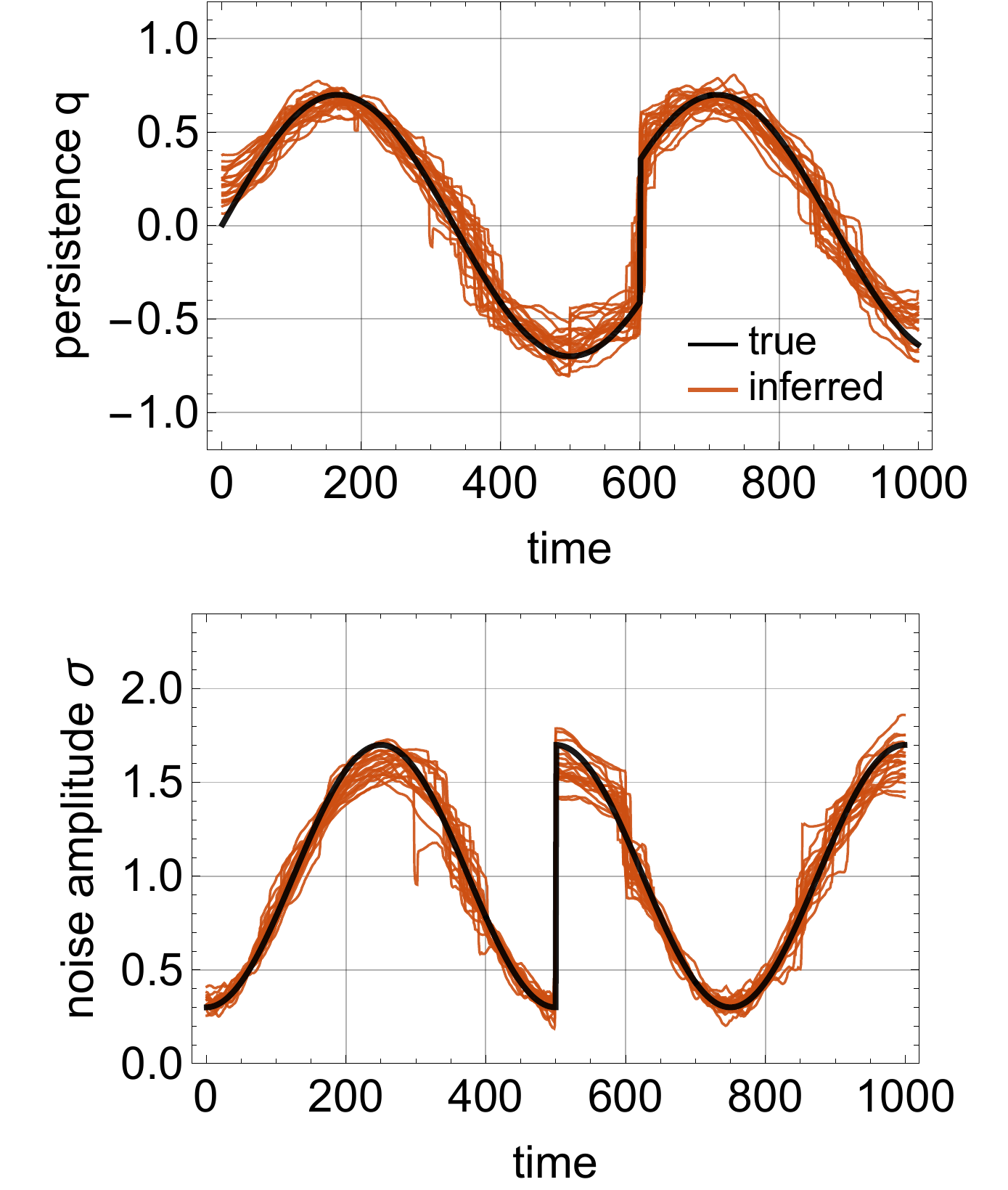}
\caption{\label{fig:sinrecon}
Inference of sinusoidal parameter changes (black). Red and gray lines show inferred parameter values of persistence (top) and noise amplitude (bottom) for $20$ realizations of the TVAR(1) process using the true parameters values.}
\end{center}
\end{figure}

\begin{figure}[h!]
\begin{center}
\includegraphics[width=0.9\columnwidth]{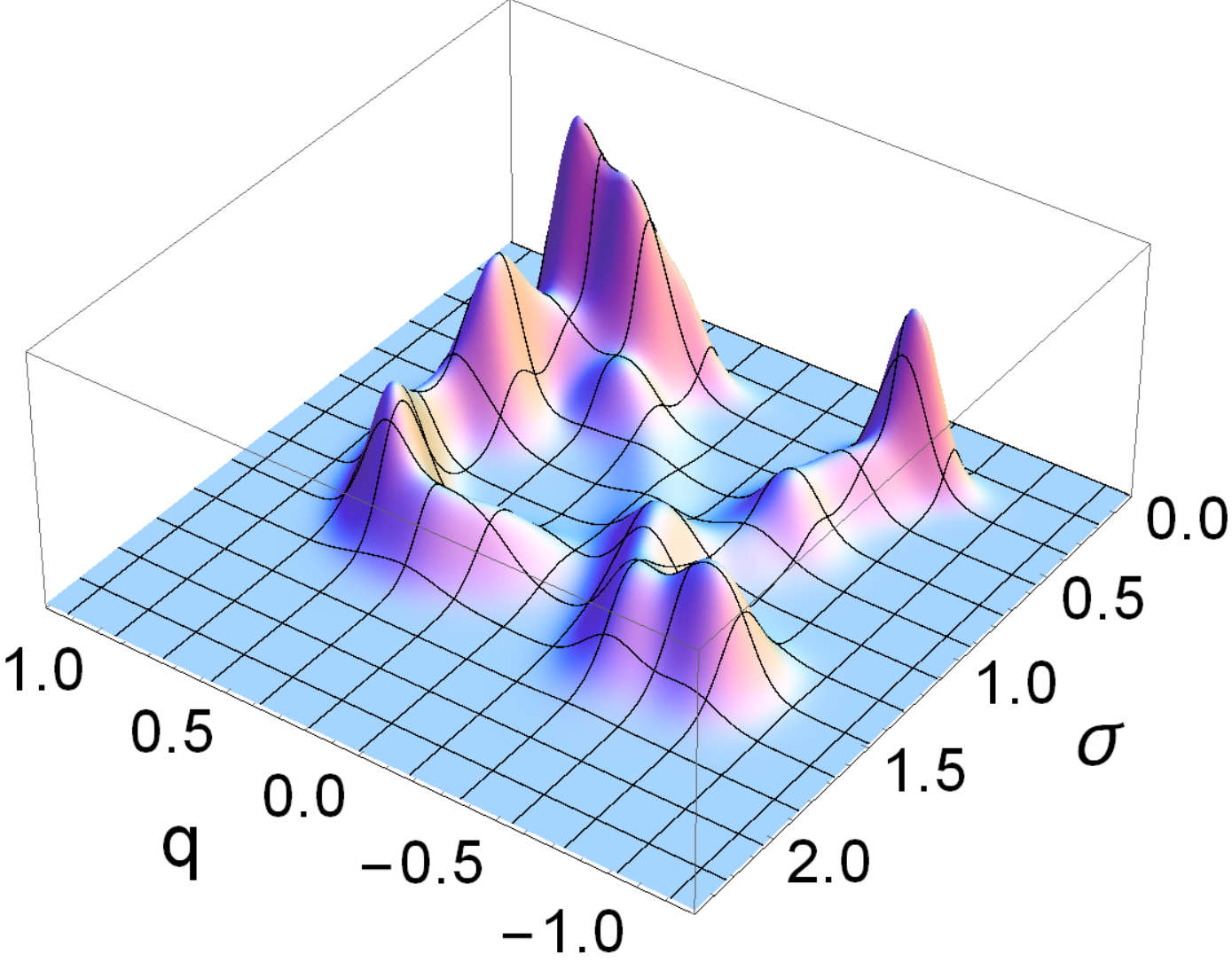}
\caption{\label{fig:sindist}
Time-averaged posterior distribution corresponding to sinusoidal parameter changes.}
\end{center}
\end{figure}

In the case presented below, we assume a sinusoidal evolution of the process parameters. A phase shift for both parameters at different times shows the response of the algorithm to an abrupt change of only one parameter. The time-varying persistence and noise amplitude, respectively, are parameterized as
\begin{align} \label{eq:sinvalues}
q_t & = \left\{
    \begin{array}{l l}
    0.7 \sin\left(\frac{3 \pi}{1000} t\right)  & \textrm{for} ~~ 0 < t \leq 600  \\
    0.7 \sin\left(\frac{\pi}{6} + \frac{3 \pi}{1000} (t-600)\right)  & \textrm{for} ~~ 600 < t \leq 1000  \\
    \end{array}
    \right.
\\
\sigma_t & = \left\{
    \begin{array}{l}
    0.8 + 0.7 \sin\left(-\frac{\pi}{2} + \frac{4 \pi}{1000} t\right)  \\
	~~~~~~~~~~~~~~~~~~~~~~~~~~~~~~~~~~~\textrm{for} ~~ 0 < t \leq 500 \\
    0.8 - 0.7 \sin\left(-\frac{\pi}{2} + \frac{4 \pi}{1000} (t-500)\right) \\
	~~~~~~~~~~~~~~~~~~~~~~~~~~~~~~~~~~~\textrm{for} ~~ 500 < t \leq 1000 \\
    \end{array}
    \right.
\end{align}
Figure \ref{fig:sinrecon} shows the true parameter values together with inferred ones, while Fig. \ref{fig:sindist} displays the time-averaged posterior distribution corresponding to the red line in Fig. \ref{fig:sinrecon}.

\subsection{Mean squared error ratio}
Given known parameter values $q_t$ and $\sigma_t$, the \emph{mean squared error} (mse) of the parameter sequences is used to assess the quality of the estimations produced by the competing methods:
\begin{align}
\operatorname{mse}\left(\{q_t\},\{\hat{q}_t\}\right) = \frac{1}{N} \sum_t \left(q_t - \hat{q}_t\right)^2 ~,
\end{align}
where $N$ denotes the length of the parameter sequence. The same formula applies to a series of noise amplitude values $\{\sigma_t\}$. The estimates of the Bayesian algorithm are denoted $\hat{q}_t^B$ and $\hat{\sigma}_t^B$. Estimations are also computed by the ML approach with a sliding window of various width $w$, denoted $\hat{q}_t^{ML(w)}$ and $\hat{\sigma}_t^{ML(w)}$.

\begin{figure}[b!]
\begin{center}
\includegraphics[width=0.9\columnwidth]{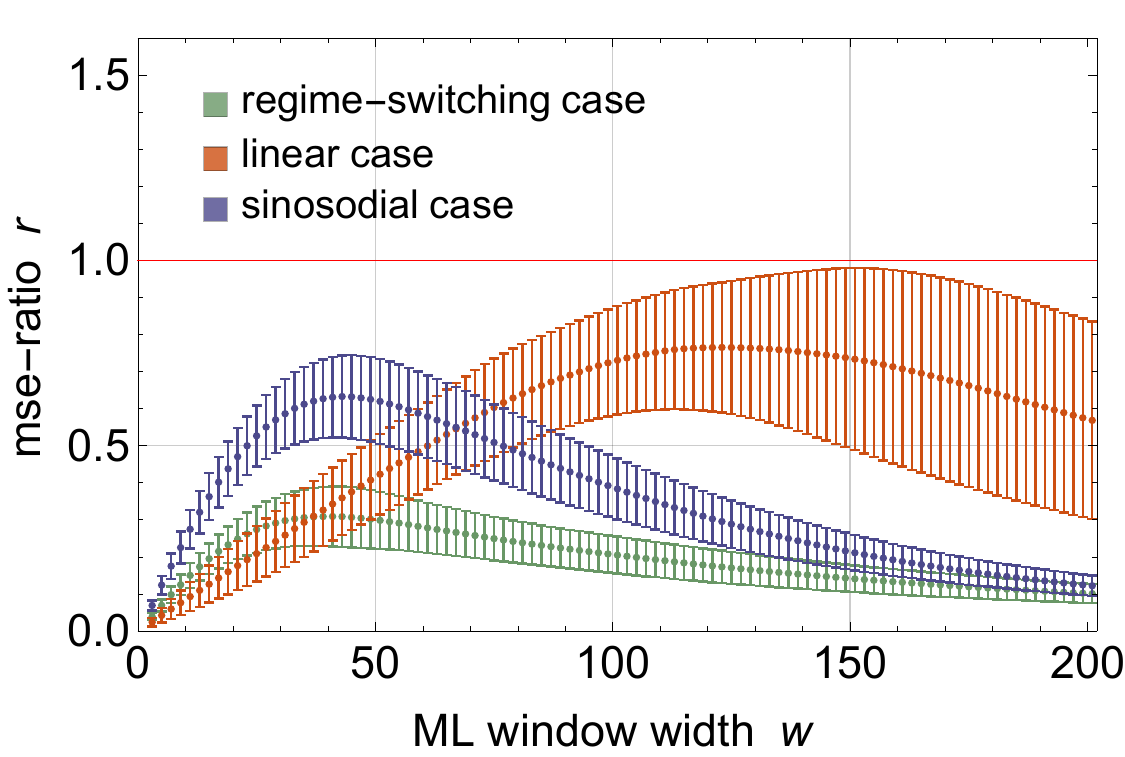}
\caption{\label{fig:comparison}
Mean squared error ratio and corresponding standard deviation plotted over the window width of the Maximum Likelihood approach for all three cases discussed above. Values of $r < 1$ denote a smaller mean squared error of the Bayesian method, compared to the sliding window.}
\end{center}
\end{figure}

In order to show that the Bayesian approach is indeed widely applicable, we compute the ML estimates using a variety of window widths $w \in \{3,5,...,201\}$, and subsequently calculate the ratio of mean squared errors of both methods, denoted $r$. Here, we add up the mean squared errors of both, persistence and noise amplitude. The ratio thus compares the Bayesian estimates to the ML estimates, depending on the chosen window width:
\begin{align}
r(\{q_t\},&\{\hat{q}_t\},\{\sigma_t\},\{\hat{\sigma}_t\}; w) = \nonumber\\
&= \frac{\operatorname{mse}(\{q_t\},\{\hat{q}_t^B\}) + \operatorname{mse}(\{\sigma_t\}, \{\hat{\sigma}_t^B\})}{\operatorname{mse}(\{q_t\},\{\hat{q}_t^{ML(w)}\}) + \operatorname{mse}(\{\sigma_t\}, \{\hat{\sigma}_t^{ML(w)}\})} ~.
\end{align}
A $\operatorname{mse}$-ratio smaller than one thus indicates a smaller estimation error for the Bayesian approach, compared to the ML approach. Figure \ref{fig:comparison} shows the mean value and standard deviation of $r$ for all three test cases, based on all $20$ trajectories of each case. On average, the Bayesian method attains a smaller mean squared error than the sliding-window approach, regardless of the chosen window width.

\section{Summary and Outlook}

In this paper, we have presented a new method to infer the time series of the hidden parameters $q_t$ and $\sigma_t$ in a TVAR(1) process, $u_t = q_t \; u_{t-1} + \sigma_t \; \epsilon_t$, from the given time series of the random variable $u_t$. We have compared the method to a Maximum Likelihood estimate of $q_t$ and $\sigma_t$ within a sliding window and demonstrated that our method is superior in reconstructing surrogate data sets for a wide range of window sizes. 

As our proposed method is based on the Bayesian framework, the possible values of the hidden parameters are for every time point $t$ described by a joint probability distribution $p(q_t,\sigma_t)$, rather than commiting to a definite point estimate. While such a point estimate can be computed directly from $p(q_t,\sigma_t)$, the width of the full distribution provides a built-in time-dependent measure of certainty for the inference process. Using a grid-based representation of $p(q_t,\sigma_t)$, no restrictions need to be imposed on the form of the distributions. In particular, multiple peaks can occur in $p(q_t,\sigma_t)$ when it is momentary uncertain whether the hidden parameters have jumped to a new pair of values. 

The time-averaged distribution $\left\langle \po(q_t,\sigma_t) \right\rangle_t$ is another interesting quantity, as it summarizes the dynamics of the hidden parameters and may be used to identify different regimes, or clusters, in parameter space. Such regimes may be distinct, i.e. separated from each other by broad borders of vanishing probability, or partially overlapping. In the latter case, inference algorithms that assume a set of discrete hidden states, such as Hidden Markov Models, often fail, while our proposed method makes no assumptions about the distribution of hidden parameters.

Finally, we would like to mention that our method could be taylored to specific cases, potentially resulting in an improved performance. In particular, we have used so far a Kernel $K$ that simultaneously accounts for slow gradual changes of the hidden parameters (part $K_2$) and for arbitrary far abrupt jumps (part $K_1$). If more detailed information were available about the temporal evolution of the hidden parameters, this could be directly incorporated into the shape of the Kernel. 

\bibliography{bibliography/biblio}

\begin{thebibliography}{10}%
\makeatletter
\providecommand \@ifxundefined [1]{%
 \@ifx{#1\undefined}
}%
\providecommand \@ifnum [1]{%
 \ifnum #1\expandafter \@firstoftwo
 \else \expandafter \@secondoftwo
 \fi
}%
\providecommand \@ifx [1]{%
 \ifx #1\expandafter \@firstoftwo
 \else \expandafter \@secondoftwo
 \fi
}%
\providecommand \natexlab [1]{#1}%
\providecommand \enquote  [1]{``#1''}%
\providecommand \bibnamefont  [1]{#1}%
\providecommand \bibfnamefont [1]{#1}%
\providecommand \citenamefont [1]{#1}%
\providecommand \href@noop [0]{\@secondoftwo}%
\providecommand \href [0]{\begingroup \@sanitize@url \@href}%
\providecommand \@href[1]{\@@startlink{#1}\@@href}%
\providecommand \@@href[1]{\endgroup#1\@@endlink}%
\providecommand \@sanitize@url [0]{\catcode `\\12\catcode `\$12\catcode
  `\&12\catcode `\#12\catcode `\^12\catcode `\_12\catcode `\%12\relax}%
\providecommand \@@startlink[1]{}%
\providecommand \@@endlink[0]{}%
\providecommand \url  [0]{\begingroup\@sanitize@url \@url }%
\providecommand \@url [1]{\endgroup\@href {#1}{\urlprefix }}%
\providecommand \urlprefix  [0]{URL }%
\providecommand \Eprint [0]{\href }%
\providecommand \doibase [0]{http://dx.doi.org/}%
\providecommand \selectlanguage [0]{\@gobble}%
\providecommand \bibinfo  [0]{\@secondoftwo}%
\providecommand \bibfield  [0]{\@secondoftwo}%
\providecommand \translation [1]{[#1]}%
\providecommand \BibitemOpen [0]{}%
\providecommand \bibitemStop [0]{}%
\providecommand \bibitemNoStop [0]{.\EOS\space}%
\providecommand \EOS [0]{\spacefactor3000\relax}%
\providecommand \BibitemShut  [1]{\csname bibitem#1\endcsname}%
\let\auto@bib@innerbib\@empty
\bibitem [{\citenamefont {Beck}\ and\ \citenamefont {Cohen}(2003)}]{Beck_2003}%
  \BibitemOpen
  \bibfield  {author} {\bibinfo {author} {\bibfnamefont {C.}~\bibnamefont
  {Beck}}\ and\ \bibinfo {author} {\bibfnamefont {E.}~\bibnamefont {Cohen}},\
  }\href {\doibase 10.1016/s0378-4371(03)00019-0} {\bibfield  {journal}
  {\bibinfo  {journal} {Physica A: Statistical Mechanics and its Applications}\
  }\textbf {\bibinfo {volume} {322}},\ \bibinfo {pages} {267–275} (\bibinfo
  {year} {2003})}\BibitemShut {NoStop}%
\bibitem [{\citenamefont {Beck}\ \emph {et~al.}(2005)\citenamefont {Beck},
  \citenamefont {Cohen},\ and\ \citenamefont {Swinney}}]{Beck_2005}%
  \BibitemOpen
  \bibfield  {author} {\bibinfo {author} {\bibfnamefont {C.}~\bibnamefont
  {Beck}}, \bibinfo {author} {\bibfnamefont {E.}~\bibnamefont {Cohen}}, \ and\
  \bibinfo {author} {\bibfnamefont {H.}~\bibnamefont {Swinney}},\ }\href
  {\doibase 10.1103/physreve.72.056133} {\bibfield  {journal} {\bibinfo
  {journal} {Phys. Rev. E}\ }\textbf {\bibinfo {volume} {72}} (\bibinfo {year}
  {2005}),\ 10.1103/physreve.72.056133}\BibitemShut {NoStop}%
\bibitem [{\citenamefont {Van~der Straeten}\ and\ \citenamefont
  {Beck}(2009)}]{Van_der_Straeten_2009}%
  \BibitemOpen
  \bibfield  {author} {\bibinfo {author} {\bibfnamefont {E.}~\bibnamefont
  {Van~der Straeten}}\ and\ \bibinfo {author} {\bibfnamefont {C.}~\bibnamefont
  {Beck}},\ }\href {\doibase 10.1103/physreve.80.036108} {\bibfield  {journal}
  {\bibinfo  {journal} {Phys. Rev. E}\ }\textbf {\bibinfo {volume} {80}}
  (\bibinfo {year} {2009}),\ 10.1103/physreve.80.036108}\BibitemShut {NoStop}%
\bibitem [{\citenamefont {Beck}(2011)}]{Beck_2011}%
  \BibitemOpen
  \bibfield  {author} {\bibinfo {author} {\bibfnamefont {C.}~\bibnamefont
  {Beck}},\ }\href {\doibase 10.1098/rsta.2010.0280} {\bibfield  {journal}
  {\bibinfo  {journal} {Philosophical Transactions of the Royal Society A:
  Mathematical, Physical and Engineering Sciences}\ }\textbf {\bibinfo {volume}
  {369}},\ \bibinfo {pages} {453–465} (\bibinfo {year} {2011})}\BibitemShut
  {NoStop}%
\bibitem [{\citenamefont {Zivot}\ and\ \citenamefont
  {Wang}(2006)}]{Zivot_2006}%
  \BibitemOpen
  \bibfield  {author} {\bibinfo {author} {\bibfnamefont {E.}~\bibnamefont
  {Zivot}}\ and\ \bibinfo {author} {\bibfnamefont {J.}~\bibnamefont {Wang}},\
  }\href {\doibase 10.1007/978-0-387-32348-0_9} {\bibfield  {journal} {\bibinfo
   {journal} {Modeling Financial Time Series with S-PLUS}\ ,\ \bibinfo {pages}
  {313–360}} (\bibinfo {year} {2006})}\BibitemShut {NoStop}%
\bibitem [{\citenamefont {Sun}\ \emph {et~al.}(2011)\citenamefont {Sun},
  \citenamefont {Qing},\ and\ \citenamefont {Wang}}]{Sun_2011}%
  \BibitemOpen
  \bibfield  {author} {\bibinfo {author} {\bibfnamefont {K.}~\bibnamefont
  {Sun}}, \bibinfo {author} {\bibfnamefont {R.}~\bibnamefont {Qing}}, \ and\
  \bibinfo {author} {\bibfnamefont {N.}~\bibnamefont {Wang}},\ }\href {\doibase
  10.1007/978-3-642-25194-8_35} {\bibfield  {journal} {\bibinfo  {journal}
  {Advances in Intelligent and Soft Computing}\ ,\ \bibinfo {pages}
  {289–296}} (\bibinfo {year} {2011})}\BibitemShut {NoStop}%
\bibitem [{\citenamefont {Hayes}(1996)}]{Hayes_1996}%
  \BibitemOpen
  \bibfield  {author} {\bibinfo {author} {\bibfnamefont {M.~H.}\ \bibnamefont
  {Hayes}},\ }in\ \href@noop {} {\emph {\bibinfo {booktitle} {Statistical
  Digital Signal Processing and Modeling}}}\ (\bibinfo  {publisher} {Wiley},\
  \bibinfo {year} {1996})\BibitemShut {NoStop}%
\bibitem [{\citenamefont {Hall}\ \emph {et~al.}(1977)\citenamefont {Hall},
  \citenamefont {Oppenheim},\ and\ \citenamefont {Willsky}}]{Hall_1977}%
  \BibitemOpen
  \bibfield  {author} {\bibinfo {author} {\bibfnamefont {M.}~\bibnamefont
  {Hall}}, \bibinfo {author} {\bibfnamefont {A.}~\bibnamefont {Oppenheim}}, \
  and\ \bibinfo {author} {\bibfnamefont {A.}~\bibnamefont {Willsky}},\ }\href
  {\doibase 10.1109/cdc.1977.271732} {\bibfield  {journal} {\bibinfo  {journal}
  {1977 IEEE Conference on Decision and Control including the 16th Symposium on
  Adaptive Processes and A Special Symposium on Fuzzy Set Theory and
  Applications}\ } (\bibinfo {year} {1977}),\
  10.1109/cdc.1977.271732}\BibitemShut {NoStop}%
\bibitem [{\citenamefont {Rabiner}(1989)}]{Rabiner_1989}%
  \BibitemOpen
  \bibfield  {author} {\bibinfo {author} {\bibfnamefont {L.}~\bibnamefont
  {Rabiner}},\ }\href {\doibase 10.1109/5.18626} {\bibfield  {journal}
  {\bibinfo  {journal} {Proceedings of the IEEE}\ }\textbf {\bibinfo {volume}
  {77}},\ \bibinfo {pages} {257–286} (\bibinfo {year} {1989})}\BibitemShut
  {NoStop}%
\bibitem [{\citenamefont {Bhanot}(1988)}]{Bhanot_1988}%
  \BibitemOpen
  \bibfield  {author} {\bibinfo {author} {\bibfnamefont {G.}~\bibnamefont
  {Bhanot}},\ }\href {\doibase 10.1088/0034-4885/51/3/003} {\bibfield
  {journal} {\bibinfo  {journal} {Reports on Progress in Physics}\ }\textbf
  {\bibinfo {volume} {51}},\ \bibinfo {pages} {429–457} (\bibinfo {year}
  {1988})}\BibitemShut {NoStop}%
\end{thebibliography}%

\end{document}